\documentclass[letterpaper]{JHEP3}

\usepackage{amsmath}

\title{Critical gravity waves}

\author{Eloy Ay\'on--Beato\\
Departamento~de~F\'{\i}sica,~CINVESTAV--IPN,~%
Apdo.~Postal~14--740,~07000,~M\'exico~D.F.,~M\'exico.\\
E-mail: \email{ayon-beato-at-fis.cinvestav.mx}}
\author{Gaston Giribet\\
The Abdus Salam International Centre for Theoretical Physics, Strada Costiera 11, 34100, Trieste, Italy.\\
Universidad de Buenos Aires and IFIBA -
CONICET, Ciudad Universitaria, Pab. I, 1428, Buenos Aires, Argentina.\\
E-mail: \email{gaston-at-ictp.it}}
\author{Mokhtar Hassa\"{\i}ne\\
Instituto de
Matem\'atica y F\'{\i}sica, Universidad de Talca, Casilla 747,
Talca, Chile.\\
Orsay Paris Laboratoire dePhysique Th\'eorique, Universit\'e Paris-Sud CNRS UMR 8627, F-1405, Orsay, France.\\
E-mail: \email{hassaine-at-inst-mat.utalca.cl}}

\abstract{ Critical Gravity in $D$ dimensions is discussed from the point of view of its exact solutions. The special features that certain 
type of solutions of higher-curvature gravity develop when one approaches the critical point of the parameter space are reviewed. In particular, a 
non-linear realization of the logarithmic modes of linearized Critical Gravity is seen to emerge as a peculiarity of the sector of anti-de Sitter wave 
solutions. Logarithmic solutions are shown to occur at a second point of the parameter space, at which the effective mass of the anti-de Sitter waves equals the 
Breitenlohner-Freedman bound. Other type of solutions with anisotropic scale invariance are also discussed and the special features 
they develop at the critical point are studied as well.

This note is the written version of the talk delivered by one of the authors
at the 13th Marcel Grossmann Meeting, held in Stockholm, Sweden, in
July 2012. The proceeding contribution is partially based on the authors' previous work \cite{nosB, nosS, nosL, nosL3D}.
}
\input{tcilatex}

\begin{document}

\section{Critical gravity}

In reference \cite{LuPope}, L\"u and Pope addressed the problem of whether a four-dimensional analogue of the three-dimensional chiral gravity 
constructed in \cite{CG} exists\footnote{See \cite{S2, S4, Grumiller, Maloney} for discussions about the three-dimensional model.}. The strategy was to consider square-curvature modifications to four-dimensional Einstein-Hilbert 
action with negative 
cosmological constant, and demand the coefficients of the action to take specific values to exclude the ghost-like excitations. While a fine tuning 
between the two different coefficients of the square-curvature action is enough to decouple the scalar excitations, demanding the cosmological constant 
to take a specific value leads the massive graviton-like excitations to become massless. As a consequence, the square-curvature theory with negative 
cosmological constant renders free of ghosts about its anti-de Sitter (AdS) background, although, as it happens in three dimensions, logarithmic modes 
that may spoil unitarity appear.

The construction of \cite{LuPope} has received the name of Critical Gravity, and it was subsequently extended to $D>4$ dimensions in reference 
\cite{Deser}; other extensions, like its supersymmetric version and its higher-curvature analogues, were discussed in \cite{susy, relaxed, cubic} and 
references thereof. Below we will discuss Critical Gravity in dimension $D=4$ and higher in detail and study special sector of its space of solutions.

An interesting aspect of Critical Gravity, and which has shown to be crucial to investigate the consistency of the theory as a toy model of quantum 
gravity \cite{logmassimo}, is the existence of spin-2 logarithmic modes \cite{LuPope}. These modes, hereafter referred to as log-modes, are analogous 
to 
those found in Topologically Massive Gravity at the chiral point\footnote{and in its analogue, the New Massive Gravity of \cite{NMG}.} \cite{Grumiller, 
Maloney}, and within the context of the four-dimensional model these were also studied in \cite{logeric, logturco, logirani}. Here, we will analyze the 
log-modes of $D$-dimensional Critical Gravity at the level of the exact solutions. Our analysis is based on our previous works \cite{nosB, nosS} and is 
close to the analysis done in 
reference \cite{logirani}. 

Let us begin by reviewing Critical Gravity in $D \geq 4$ dimensions.

\subsection{Critical Gravity in $D=4$ dimensions}

Critical Gravity in $D=4$ is defined by the action \cite{LuPope}%
\begin{equation}
S[g_{\mu \nu }]=\int {d}^{4}x\sqrt{-g}\left( R-2\Lambda +\beta _{1}{R}%
^{2}+\beta _{2}{R}_{\alpha \beta }{R}^{\alpha \beta }\right) .  \label{S}
\end{equation}%
with its coupling constants restricted to obey the relation%
\begin{equation}
\beta _{2}=-3\beta _{1}  \label{cond1}
\end{equation}%
together with%
\begin{equation}
\beta_1 =-\frac{1}{2\Lambda }.  \label{cond2}
\end{equation}

The relation (\ref{cond1}) between the coupling constants $\beta_i$ implies that the coefficient of the $\square R$ contribution to the trace of the 
equations of motion vanishes; see (\ref{eq:squareGrav}) below. Then, at this point of the parameter space the theory looses its scalar mode. On the 
other hand, while 
condition 
(\ref{cond1}) is 
needed to decouple the spin-0 excitation, condition (\ref{cond2}) is needed 
for the spin-2 excitations of the
theory to become massless. The latter condition may, however, be relaxed
and, following \cite{relaxed}, the model with the cosmological constant taking values
in the range $0<\beta _{1}\leq -1/(2\Lambda )$ may be considered. The model of \cite{LuPope} then appears as the upper
bound of this window.

Because of the Gauss-Bonnet theorem, we know that in four dimensions the Kretschmann scalar $%
R_{\alpha \beta \gamma \delta }R^{\alpha \beta \gamma \delta }$ can be
written in terms of the square-curvature terms appearing in (\ref{S}), up to
a total derivative. That is, action (\ref{S}) is the most general quadratic
action in three and four dimensions. As said, Critical Gravity is defined by
considering (\ref{S}) with (\ref{cond1})-(\ref{cond2}), and in many ways it
is analogous to the three-dimensional chiral gravity introduced in \cite{CG}
and its generalization that includes the model proposed in \cite{NMG}. The main goal of the present work is to emphasize this analogy by
stablishing a parallelism between our previous work \cite{nosB} and the $%
D\geq 4$ critical models.

\subsection{Extension to $D>4$ dimensions}

Unlike the three- and the four-dimensional cases, in dimension $D>4$ three
different invariants have to be used to write the most general quadratic
action; namely 
\begin{equation}
S[g_{\mu \nu }]=\int {d}^{D}x\sqrt{-g}\left( R-2\Lambda +\beta _{1}{R}%
^{2}+\beta _{2}{R}_{\alpha \beta }{R}^{\alpha \beta }+\beta _{3}{R}_{\alpha
\beta \mu \nu }{R}^{\alpha \beta \mu \nu }\right) .  \label{eq:Squad}
\end{equation}

Critical Gravity in $D$-dimensions corresponds to considering this action
with the following tuning of the coupling constants 
\begin{equation}
\beta _{1}=\frac{2\beta _{3}}{(D-1)(D-2)},\qquad \beta _{2}=-\frac{4\beta
_{3}}{(D-2)},  \label{torbellino}
\end{equation}%
and also%
\begin{equation}
\Lambda =\frac{(D-1)(D-2)}{8\beta _{3}(D-3)}.  \label{Lambda}
\end{equation}

The linear combination (\ref{torbellino}) of the quadratic terms in (\ref{eq:Squad})
is such that, up to a total derivative, it coincides with $\sqrt{-g}%
C_{\alpha \beta \gamma \delta }C^{\alpha \beta \gamma \delta }$, where $%
C_{\alpha \beta \gamma \delta }$ is the Weyl tensor; in four dimensions,
this modification is conformally invariant. 

The strategy here will be first considering the general quadratic gravity action and then
see what special features appear at the critical point. In turn, let us begin by writing
the field equations that correspond to vary (\ref{eq:Squad}) with respect to the metric,
\begin{eqnarray}
&&R_{\mu \nu }-\frac{1}{2}Rg_{\mu \nu }+\Lambda {g}_{\mu \nu }{}-\frac{1}{2}%
\left( \beta _{1}{R}^{2}+\beta _{2}{R}_{\alpha \beta }{R}^{\alpha \beta
}+\beta _{3}{R}_{\alpha \beta \gamma \delta }{R}^{\alpha \beta \gamma \delta
}\right) g_{\mu \nu }  \notag \\
&&{}+2\beta _{3}R_{\mu \gamma \alpha \beta }R_{\nu }^{~\gamma \alpha \beta
}+2\left( \beta _{2}+2\beta _{3}\right) R_{\mu \alpha \nu \beta }R^{\alpha
\beta }-4\beta _{3}R_{\mu \alpha }R_{\nu }^{~\alpha }+2\beta _{1}RR_{\mu \nu
}  \notag \\
&&+\left( \beta _{2}+4\beta _{3}\right) \square {R}_{\mu \nu }+\frac{1}{2}%
\left( 4\beta _{1}+\beta _{2}\right) g_{\mu \nu }\square {R}-\left( 2\beta
_{1}+\beta _{2}+2\beta _{3}\right) \nabla _{\mu }\nabla _{\nu }{R}=0.\qquad ~
\label{eq:squareGrav}
\end{eqnarray}

We already notice from this that mode $\square {R}$ actually decouples as it does not
appear in the trace of the field equations. 

Before deriving different classes of solutions to (\ref{eq:squareGrav}), let us first fix the
cosmological constant $\Lambda $ such that the AdS$_{D}$ spacetime of radius 
$l$ is a solution of the equations (\ref{eq:squareGrav}). Then, we find the
following constraint between the cosmological constant $\Lambda $, the AdS$%
_{D}$ radius $l$, and the coupling constants $\beta _{i}$ 
\begin{equation}
\Lambda =-\frac{(D-1)(D-2)}{2l^{2}}+\frac{(D-1)(D-4)}{2l^{4}}\left[
(D-1)\left( D\beta _{1}+\beta _{2}\right) +2\beta _{3}\right] .
\label{eq:lambda}
\end{equation}%
From this, we notice that only in four dimensions the cosmological constant
is related to the AdS radius in the usual way, without involving the
couplings $\beta _{i}$. For $D\neq 4$ the value of $l$ is uniquely given in terms of $\Lambda $ for very special values of $\beta_i$; see blow. In the 
generic case, there are two maximally symmetric backgrounds provided
the effective cosmological constant $-l^{-2}$ may take two different values. In Critical Gravity (\ref{torbellino})-(\ref{Lambda}), on the contrary, the second term on the right hand side of
(\ref{eq:lambda}) vanishes and thus the vacuum is unique.

\section{Anti-de Sitter waves \label{sec:SSiklos}}

Here, we will explore exact solutions to Critical Gravity in $D$ dimensions.
We consider an ansatz of the following form 
\begin{equation}
d{s}^{2}=\frac{l^{2}}{r^{2}}\left[ -\left( 1+2{h} \right) dt^{2}+2dtd\xi
+dr^{2}+d\vec{x}^{2}\right] ,  \label{eq:ansatz}
\end{equation}%
where ${h} $ is a metric function that does not depend on the lightlike
coordinate $\xi $. Here, $d\vec{x}^{2}$ refers to the Euclidean metric on
the $\mathbb{R}^{D-3}$ plane. We will consider deformations of the universal
covering of AdS$_{D}$, so the timelike coordinate takes values $t\in \mathbb{%
R}$, as well as the lightlike coordinate $\xi \in \mathbb{R}$. The radial
coordinate takes values $r\in \mathbb{R}_{\geq 0}$. For ${h} =const$, we
recover the metric of AdS$_{D}$ space in Poincar\'{e} coordinates, where the
boundary of AdS$_{D}$ is located at $r=0$. Metric (\ref{eq:ansatz})
corresponds to a special class of the so-called Siklos spacetimes \cite{Siklos:1985}; and in the general case, solutions (\ref{eq:ansatz}) can be
though of as describing exact gravitational waves propagating on the AdS$_{D}
$ spacetime \cite{Podolsky:1997ik} (hereafter referred to as AdS-waves).
Solutions (\ref{eq:ansatz}) are, indeed, a special case of a more
general family of solutions studied in \cite{Garcia:1981}.

Here, we will be mostly concerned with AdS-wave solutions of the type (\ref%
{eq:ansatz}) in the case that ${h} $ only depends on the radial coordinate $%
r$. Nevertheless, it is worth mentioning that most of the formulae below
remains valid when the profile of the wave corresponds to a more general
Siklos solution with ${h} ={h} (t,r,\vec{x})$. We use the null geodesic
vector $k^{\mu }\partial _{\mu }=(r/l)\partial _{\xi }$ that allows us to
interpret these backgrounds as generalized Kerr-Schild transformations of AdS%
$_{D}$;\ namely 
\begin{equation}
g_{\mu \nu }=g_{\mu \nu }^{\mathrm{AdS}}-2{h} \ k_{\mu }k_{\nu }.
\label{eq:K-S}
\end{equation}%
where $g_{\mu \nu }^{\mathrm{AdS}}$ is the metric of AdS$_{D}$, and recall $k_{\mu
}k^{\mu }=0$.

The Ricci tensor for a metric like (\ref{eq:K-S}) takes the form 
\begin{equation}
R_{\mu \nu }=-\frac{(D-1)}{l^{2}}g_{\mu \nu }+k_{\mu }k_{\nu }\square {{h} }%
,  \label{eq:Ricci}
\end{equation}%
and it yields constant scalar curvature $R=-D(D-1)/l^{2}$, exactly the same
as for AdS$_{D}$ space. It also gives the squared-curvature combinations%
\begin{eqnarray}
R_{\mu \alpha }R_{\nu }^{~\alpha } &=&\frac{(D-1)^{2}}{l^{4}}g_{\mu \nu }-%
\frac{2(D-1)}{l^{2}}k_{\mu }k_{\nu }\square {{h} },
\label{eq:Riemann*Ricci} \\
R_{\mu \alpha \nu \beta }R^{\alpha \beta } &=&\frac{(D-1)^{2}}{l^{4}}g_{\mu
\nu }-\frac{(D-2)}{l^{2}}k_{\mu }k_{\nu }\square {{h} },\qquad 
\label{eq:Ricci*Ricci} \\
R_{\mu \gamma \alpha \beta }R_{\nu }^{~\gamma \alpha \beta } &=&\frac{2(D-1)%
}{l^{4}}g_{\mu \nu }-\frac{4}{l^{2}}k_{\mu }k_{\nu }\square {{h} }.
\label{eq:Riemann*Riemann}
\end{eqnarray}

Using the expression for the Ricci tensor (\ref{eq:Ricci}), together with
the null and geodesic properties of $k^{\mu }$, one finds that all the
invariants constructed from the Riemann tensor turn out to coincide with
those of AdS$_{D}$ spacetime. Besides, one finds that the only contribution
to the equations of motion that contains derivatives of the Riemann tensor
is 
\begin{equation}
\square {R}_{\mu \nu }=k_{\mu }k_{\nu }\square \left( \square -\frac{2}{l^{2}%
}\right) {h} .  \label{eq:Box_Ricci}
\end{equation}

Then, taking into account that $\Lambda $ is given by (\ref{eq:lambda}), one finally
finds that the equations of motion (\ref{eq:squareGrav}) reads
\begin{equation}
(\beta _{2}+4\beta _{3})\ k_{\mu } k_{\nu } \ (\square -m^{2})\square {h} =0.  \label{GGGG}
\end{equation}
where $m^{2}$ is given by 
\begin{equation}
m^{2}=\frac{2(D-1)(D\beta _{1}+\beta _{2})-4(D-4)\beta _{3}-l^{2}}{%
l^{2}(\beta _{2}+4\beta _{3})}.  \label{unamasa}
\end{equation}%

We will be specially interested in solutions that tend to AdS$_{D}$
spacetime close to the boundary. In turn, as always when
dealing with asymptotically AdS$_{D}$ spaces, a crucial question is that
about the next-to-leading dependence in large distances (i.e. large $1/r$)
expansion. Assuming that the function ${h} $ only depends on the radial
coordinate $r$, the equations of motion reduces to the single homogeneous Euler differential equation 
\begin{eqnarray}
&&(\beta _{2}+4\beta _{3})\left[ r^{2}{h} ^{\prime \prime \prime \prime
}-2(D-4)r{h} ^{\prime \prime \prime }\right] +  \notag \\
&&+\big[l^{2}-2D(D-1)\beta _{1}+(D-2)(D-8)\beta _{2}+4(D-2)(D-5)\beta _{3}%
\big]{h} ^{\prime \prime }-  \notag \\
&&-(D-2)\left[ l^{2}-2D(D-1)\beta _{1}-(3D-4)\beta _{2}-8\beta _{3}\right] 
\frac{1}{r}{h} ^{\prime }=0,  \label{eq:SGAdSwave}
\end{eqnarray}%
where the prime stands for the derivative with respect to $r$, $f^{\prime
}=\partial _{r}f$, and where we used that $\square f=\left[
r^{2}f^{\prime \prime }-(D-2)rf^{\prime }\right] /l^{2}$. 

Equation (\ref%
{eq:SGAdSwave})\ actually corresponds to the $tt$ component of the equations
of motion, which is the only non-trivial one. In what follows, we provide a
detailed analysis of the equation (\ref{eq:SGAdSwave}) and its solutions; in
particular, we will see that non-trivial asymptotically AdS$_{D}$ solutions
exist.

\subsection{Massless waves of the second order sector \label{subsec:so}}

Let us begin by noticing that, although in the generic case (\ref%
{eq:SGAdSwave}) is a fourth order linear differential equation, in the particular
case $\beta _{2}=-4\beta _{3}$ the first line in (\ref{eq:SGAdSwave})
dissapears and then the equation renders of second order, yielding 
\begin{equation}
\left[ l^{2}-2D(D-1)\beta _{1}+12(D-2)\beta _{3}\right] \ \square {h} =0;
\label{eq:secord}
\end{equation}%
notice that $\square $, which corresponds to the d'Alambertian operator
associated to metric (\ref{eq:ansatz}), does not to depend on the function $%
{h} $, so that in particular it coincides with the d'Alambetian operator in
AdS$_{D}$ spacetime.

If in addition to $\beta _{2}=-4\beta _{3}$ we considered $\beta
_{1}=[l^{2}+12(D-2)\beta _{3}]/[2D(D-1)]$ then full degeneracy would occur
and the field equations would then be satisfied for any profile function $%
{h} $. This degenerate point of the space of parameter is also interesting
and, for instance, in $D=5$ dimensions it includes the five-dimensional
Chern-Simons gravity theory for the group SO($2,4$), for which this type of degeneracy is known to
exist. On the contrary, for $\beta _{1}\neq \lbrack l^{2}+12(D-2)\beta
_{3}]/[2D(D-1)]$, the resulting equation is strictly of second order, as in
the case of Lovelock theory \cite{Lovelock}. In fact, Lovelock theory is a
particular case of the former, which corresponds to having $\beta
_{2}=-4\beta _{3}$ together with $\beta _{1}=\beta _{3}$.

In general, in the second order case $\beta _{2}=-4\beta _{3}$ the solution
is given by 
\begin{equation}
{h} (r)=c_{0}\ r^{D-1},  \label{eq:GR}
\end{equation}%
up to an arbitrary constant that can always be removed by a change of
coordinates. Not surprisingly, (\ref{eq:GR}) coincides with a solution of
General Relativity (GR) with negative cosmological constant, i.e. the theory
that corresponds to $\beta _{i}=0$. This solution is, indeed, the usual GR
exact massless scalar mode that propagates on AdS$_{D}$ spacetime, and the field
equation (\ref{eq:secord}) in that case reduces to the wave equation $%
\square {{h} }=0$, with $\square ${\ being the d'Alambertian operator of AdS%
}$_{D}$. 

Generically, a solution like (\ref{eq:ansatz}) is physically interpreted as an exact massive gravitational wave propagating
on the AdS$_{D}$ spacetime also in the cases $\beta _{i}\neq 0$. In fact, for generic values of $\beta _{2}\neq -4\beta _{3},$ one
finds that at least one the set of solutions to (\ref{GGGG}) obeys the
Klein-Gordon wave function $(\square -m^{2}){{h} }=0$ with an effective
mass parameter $m$ being given by (\ref{unamasa}). We will explore these
solutions in detail below.

\subsection{Massive waves in anti-de Sitter spacetime\label{subsec:Sis}}

For $\beta _{2}\neq -4\beta _{3}$, field equation (\ref{eq:SGAdSwave}) is a
fourth-order Euler differential equation which, in the generic case, admits
linearly independent solutions of the power-law form ${h} \sim r^{\alpha }$%
, with the exponents $\alpha $ being the roots of the fourth-degree
polynomial 
\begin{equation}
\alpha (\alpha -D+1)\left[ \left( \alpha -\frac{D-1}{2}\right) ^{2}-\frac{%
(D-1)^{2}}{4}-\frac{l^{2}-2(D-1)(D\beta _{1}+\beta _{2})+4(D-4)\beta _{3}}{%
\beta _{2}+4\beta _{3}}\right] =0.  \label{eq:charpol}
\end{equation}%
Since the constant solution, i.e.\ $\alpha =0$, can be removed by coordinate
transformations, the general solution results to be given by 
\begin{equation}
{h} (r)=c_{0}\ r^{D-1}+c_{+}\ r^{\alpha _{+}}+c_{-}\ r^{\alpha _{-}},
\label{HJK}
\end{equation}%
where 
\begin{equation}
\alpha _{\pm }=\frac{D-1}{2}\pm \sqrt{\frac{(D-1)^{2}}{4}+\frac{%
2(D-1)(D\beta _{1}+\beta _{2})-4(D-4)\beta _{3}-l^{2}}{\beta _{2}+4\beta _{3}%
}},  \label{eq:alpha_pm}
\end{equation}%
and where $c_{0}$ and $c_{\pm }$ are arbitrary integration constants.

The type of solutions (\ref{HJK}) includes cases which are interesting for
physics. For instance, for $c_{0}=c_{+}=0$ it includes geometries whose
isometry group is the so-called Schr\"{o}dinger group, namely solutions of the form
\begin{equation}
d{s}^{2}=\frac{l^{2}}{r^{2}}\left( -\frac{1}{r^{2z-2}} dt^{2}+2dtd\xi
+dr^{2}+d\vec{x}^{2}\right) , \label{dosdieciseis}
\end{equation}
where $\alpha _- = 2(1-z)$. These geometries 
have recently been considered as holographic gravity duals of non-relativistic systems \cite{Son,BalasubramanianMcGreevy} with Galilean invariance and 
anisotropic scaling symmetry under $(t,\xi
,r,x)\rightarrow (\lambda ^{2z}t,\lambda ^{4-2z}\xi ,\lambda ^{2}r,\lambda
^{2}x)$. 

Among solutions (\ref{HJK}) there are
also solutions that asymptote AdS$_{D}$ spacetime at large distance. To see
this, let us define the new timelike coordinate $\hat{t}\equiv t-\xi $
and the radial coordinate $\hat{r}\equiv l^{2}/r$, so that the near
boundary region of AdS$_{D}$ corresponds to large $\hat{r}$. In these
coordinates, metric (\ref{eq:ansatz}) takes the form
\begin{equation}
d{s}^{2}=\frac{\hat{r}^{2}}{l^{2}}(-d\hat{t}^{2}+d\xi ^{2}+d\vec{x}%
^{2})+\frac{l^{2}}{\hat{r}^{2}}dr^{2}-\hat{{h}}(\hat{r})(d\hat{t}+d\xi
)^{2}, \label{dosdiecisiete}
\end{equation}%
where the function $\hat{{h}}(\hat{r})=2\hat{r}^{2}{h} (1/\hat{r}%
)/l^{2} $ represents the perturbation of AdS$_{D}$ spacetime. Asymptotically
AdS$_{D}$ solutions then correspond to solutions with $\alpha _{+}\geq 2$,
for which%
\begin{equation*}
g_{\hat{t}\hat{t}}=-\frac{\hat{r}^{2}}{l^{2}}+\mathcal{O}%
(1),\qquad g_{\hat{t}\xi }=-\frac{\hat{r}^{2}}{l^{2}}+\mathcal{O}%
(1),\qquad g_{\xi \xi }=-\frac{\hat{r}^{2}}{l^{2}}+\mathcal{O}(1),
\end{equation*}%
where $\mathcal{O}(1)$ stands for terms that either do not depend on $%
\hat{r}$ or whose $\hat{r}$-dependences decay near the boundary.
Relaxed asymptotic conditions involving logarithmic dependences in the
next-to-leading order in the $1/\hat{r}$ expansion will be discussed
below.

As mentioned, solutions (\ref{HJK}) generically satisfy the wave equation%
\footnote{%
Strictly speaking, this is true for generic $\alpha _{\pm }$, with $\alpha
_{+}\neq \alpha _{-}$ and $\alpha _{\pm }\neq D-1$; see below.} $(\square
-m^{2}){{h} }=0$ with $m$ given by (\ref{unamasa}). More precisely,
solutions (\ref{HJK}) describe the superposition of three scalar modes: the
massless mode of GR and two massive modes generated by the square-curvature
terms, the latter having the same effective mass (\ref{unamasa}). It is
worth pointing out that the solution (\ref{HJK}) is valid only when the
roots (\ref{eq:alpha_pm}) are real. This in turn constrains the mass
parameter to obey strictly the Breitenlohner-Freedman (BF)\ bound of the AdS$%
_{D}$ in which the mode propagates \cite{BF}; namely 
\begin{equation}
m^{2}>m_{\mathrm{BF}}^{2}\equiv -\frac{(D-1)^{2}}{4l^{2}}.  \label{eq:BFa}
\end{equation}

\section{Logarithmic Gravity in $D$ dimensions}

We now turn to analyze the cases for which the roots $\alpha _{\pm }$ of the
polynomial (\ref{eq:charpol}) degenerate and two would-be-independent
solutions to the differential equation coalesce. As usual, when this type of
multiplicity in the roots takes place, the power-law dependence does not
represent the more general solutions to Eq.~(\ref{eq:SGAdSwave}), and
additional behaviors, typically involving logarithmic dependences, appear.
The existence of such \emph{exact\/} logarithmic behaviors is totally
analogous to what happens in three-dimensional massive gravity; see for
instance \cite{AyonBeato:2004fq,AyonBeato:2005qq,AyonBeato:2005bm}; see also \cite{nosB}.

\subsection{The Logarithmic sectors I:\ The Breitenlohner-Freedman point.}

The first source of multiplicity appears when the roots (\ref{eq:alpha_pm})
become one single root, namely when $\alpha _{+}=\alpha _{-}$. This happens
when 
\begin{equation}
\beta _{2}=\frac{4\left[ l^{2}-2D(D-1)\beta _{1}-(D^{2}-6D+17)\beta _{3}%
\right] }{(D+7)(D-1)},.  \label{eq:mul_221}
\end{equation}%
In this resonant case the mass $m^{2}$ tends to the BF bound $m_{\mathrm{BF}%
}^{2}$ and the general solution turns to be 
\begin{equation}
{h} (r)=c_{0}\ r^{D-1}+c_{1}\ r^{\frac{D-1}{2}}\log ({r)}+c_{2}\ r^{\frac{%
D-1}{2}},  \label{eq:Fmul_221}
\end{equation}%
up to an additive constant. It is worth noticing that, while in the case of
solutions like (\ref{eq:Fmul_221}) with $c_{2}\neq 0,$ $c_{1}=0$ the profile
function ${h} $ obeys the Klein-Gordon equation 
\begin{equation}
(\square -m_{\mathrm{BF}}^{2}){{h} }=0\quad \quad \text{with}\quad \quad m_{%
\mathrm{BF}}^{2}\equiv -\frac{(D-1)^{2}}{4l^{2}},  \label{eq:BF}
\end{equation}%
in the case of solutions with $c_{1}\neq 0$ we find 
\begin{equation}
(\square -m_{\mathrm{BF}}^{2})\square {{h} }=0,
\end{equation}%
though ${h} $ does not satisfy the Klein-Gordon equation for any value of $m
$. This is similar to what was observed in three-dimensional massive gravity \cite{nosB}.

Now, let us turn to the case we are interested in, namely the case of
Critical Gravity.

\subsection{The Logarithmic sectors II:\ The Critical Gravity point.}

The second case in which multiplicity of the roots $\alpha _{\pm }$ occurs
is when one of the two generic roots (\ref{eq:alpha_pm}) either vanishes or
takes the value $D-1$. Actually, it is easy to check that these two
possibilities occur simultaneously, namely $\alpha _{-}=0$ precisely when $%
\alpha _{+}=\alpha _{+}^{\text{(GR)}}=D-1$. In this case, the coupling
constants obey 
\begin{equation}
\beta _{2}=\frac{l^{2}-2D(D-1)\beta _{1}+4(D-4)\beta _{3}}{2(D-1)},
\label{eq:sdm}
\end{equation}%
and this is actually what happens at the critical point (\ref{torbellino})-(%
\ref{Lambda}). The solution with simultaneous double multiplicity is then of
the form 
\begin{equation}
{h} (r)=c_{0}\ r^{D-1}+c_{1}r^{D-1}\log {(r)}+c_{2}\log {(r)}.  \label{eq:Fsdm}
\end{equation}
That is, Critical Gravity in $D$ dimensions admits logarithmic modes (\ref%
{eq:Fsdm}) as exact solutions, similarly as it happens in three-dimensional
massive gravity \cite{nosB}. Besides, as it happens with the BF\ logarithmic modes discussed in the
previous subsection, only solutions (\ref{eq:Fsdm}) having $c_{1}=c_{2}=0$
obey the massless wave equation $\square {{h} }=0$. The logarithmic modes,
on the contrary, obey the fourth-order equation%
\begin{equation}
\square ^{2}{{h} }=0\quad \quad \text{although}\quad \quad \square {{h} }%
\neq 0.
\end{equation}
Again, this is totally analogous to what happens with the linearized
log-modes of \cite{LuPope}. In fact, we interpret logarithmic solutions (\ref{eq:Fsdm}) as the non-linear realization of the former. 
To see this it is convenient to consider the basis
\begin{equation}
h^{GR}_{\pm} =  \frac{(D-1)}{l^2} (r^{D-1}\mp 1), \ \ \ \ \ \ h^{Log }_{\pm} =  (r^{D-1}\pm 1) \log (r),  
\end{equation}
that is $h^{Log }_{\pm} \propto h^{GR}_{\mp} \log (r)$, in which the d'Alambertian operator takes the following Jordan's cell form
\begin{equation}
\square h_{\pm}^{Log } (r) = h_{\pm }^{GR} (r), \ \ \ \ \ \ \  \square h_{\pm}^{GR} (r) = 0.   
\end{equation}

The question remains as to whether Log-modes can be consistently decoupled by an appropriate choice of boundary conditions.
As we learned from the three-dimensional case \cite{Maloney},
the question about the consistency of a tructation of the theory may be a subtle one.

\section{Other sectors of the space of solutions}

The AdS-waves is not the only sector of exact solutions of higher-curvature gravity that exhibits peculiarities at the critical point. In fact, the 
special character of the critical point can also be observed by
investigating other sectors of the space of solutions. For instance, let us consider the so-called
Lifshitz metrics in $D=4$ dimensions: Consider%
\begin{equation}
d{s}^{2}=-\frac{l^{2z}}{r^{2z}}d\hat{t}^{2}+\frac{l^{2}}{r^{2}}%
(dr^{2}+d\xi ^{2}+dx^{2}),  \label{Lifshitz}
\end{equation}%
where now the coordinate $\xi $ plays the same role as $x$, c.f. (\ref{dosdieciseis}). Metric (\ref{Lifshitz})
presents anisotropic scale invariance, as the scaling transformation $(%
\hat{t},r, \xi , x) \to (\lambda^{2z}\hat{t},\lambda^{2}r,
\lambda^{2}\xi , \lambda^{2}x)$ represents a symmetry of it; this is why
these spaces have recently been considered in the context of
non-relativistic holography as well \cite{Kachru}. Metric (\ref{Lifshitz})
reduces to AdS$_{D} $ spacetime when $z=1$;\ that is, it reduces to the case 
${h} =const$ of (\ref{eq:ansatz}).

It was shown in \cite{nosL} that metric (\ref{Lifshitz}) is also a solution
of the $D=4$ theory (\ref{S}) provided the following relations between the
parameters $l,z$ and the coupling constants are obeyed 
\begin{equation}
\beta _{2}=-3\beta _{1}=\frac{3l^{2}}{2z(z-4)},
\end{equation}
\begin{equation}
\Lambda =-\frac{z^{2}+2z+3}{2l^{2}} .
\end{equation}

That is, Lifshitz metric with arbitrary dynamical exponent $z$ is a solution
of $D=4$ square-curvature gravity with the precise relation between $\beta
_{1}$ and $\beta _{2}$ that holds in Critical Gravity, namely (\ref{cond1}).
Nevertheless, when in addition to (\ref{cond1}) one imposes the restriction (%
\ref{cond2}) then one only finds $z=1$, so that, at the critical point
the Lifshitz solutions flow to the AdS$_{D}$ spacetime. The same happens with the Schr\"{o}dinger invariant backgrounds (\ref{dosdieciseis}), which flow to the isotropic value $z=1$ when the critical point is approached. This phenomenon was also observed in three dimensions \cite{nosL3D}.

\acknowledgments

The work of G.G. was supported by grants PIP, NSF-CONICET, PICT, and UBACyT from CONICET, ANPCyT, and UBA.The work of M.H. was partially supported by grant 1090368 from
FONDECYT and by grant ACT 06 (Anillo  Lattice and Symmetry). Discussions with Alan Garbarz, Andr\'{e}s Goya, and Guill\'{e}m P\'{e}rez-Nadal are acknowledged.
G.G. specially thanks the organizers of the 13th Marcel Grossmann Meeting on General Relativity.



\end{document}